\documentclass[journal]{IEEEtran}
\usepackage{amsmath,amssymb,amsfonts}
\usepackage{algorithm}
\usepackage{algorithmic}
\usepackage{graphicx}
\usepackage{textcomp}
\usepackage{xcolor}
\usepackage{url}
\usepackage{hyperref}
\usepackage{cite}
\usepackage{booktabs}
\usepackage{braket}
\usepackage{float}

\DeclareMathOperator*{\argmax}{arg\,max}

\graphicspath{{figures/}}

\begin{document}

\title{Quantum Key Distribution Secured Federated Learning for Channel Estimation and Radar Spectrum Sensing in 6G Networks}

\author{%
Ferhat Ozgur Catak\IEEEauthorrefmark{1},
Murat Kuzlu\IEEEauthorrefmark{2},
Jungwon Seo\IEEEauthorrefmark{1},
and Umit Cali\IEEEauthorrefmark{3}\\
\IEEEauthorblockA{\IEEEauthorrefmark{1}University of Stavanger (UiS), Stavanger, Norway}
\IEEEauthorblockA{\IEEEauthorrefmark{2}Old Dominion University, Norfolk, VA, USA}
\IEEEauthorblockA{\IEEEauthorrefmark{3}University of York, York, UK}
\IEEEauthorblockA{Email: f.ozgur.catak@uis.no}
}

\maketitle

\begin{abstract}
This paper presents a federated learning framework secured by quantum key distribution (QKD) for wireless channel estimation and radar spectrum sensing in the next generation networks (NextG or Beyond 6G). A BB84-style protocol abstraction and pairwise additive masking are utilized to train clients' local models (CNN for channel estimation, U-Net for radar segmentation) and upload only masked model updates. The server aggregates without observing plain parameters; an eavesdropper without QKD keys cannot recover individual updates. Experiments show that secure FL achieves NMSE of 0.216 for channel estimation and 92.1\% accuracy with 0.72 mIoU for radar sensing. When an eavesdropper is present, QBER rises to $\sim$25\% and all rounds abort as intended; reconstruction error remains below $10^{-5}$, confirming correct aggregation.
\end{abstract}



\section{Introduction}
\label{sec:intro}

In recent years, Next-generation networks (NextG or Beyond 6G) have been enhanced with advanced communication and computing technologies to meet the requirements of high data-speed and low latency, i.e., up to Tbps and less than milliseconds, for a wide range of applications, from autonomous vehicles to industry 5.0 and advanced manufacturing, telehealth and remote surgery, online education, immersive virtual reality (VR) and augmented reality (AR) \cite{jain2025recent, ogenyi2025comprehensive, kuzlu2025ai}. NextG networks are envisioned to integrate sensing and communication technologies, enabling joint radar and wireless operations in shared spectrum. Fortunately, key innovative advances such as 
massive multiple-input multiple-output (MIMO), spectrum sensing, beamforming,  Intelligent Reflecting Surfaces (IRS), network slicing, and Terahertz Communication \cite{sun2023massive, othman2025key, abba2025iot, catak2022security} address those requirements, along with the integration of artificial intelligence/machine learning approaches to improve overall system performance and
efficiency, optimizing network management and traffic enhancement techniques, enhancing security, and dynamically allocating resources. However, utilizing AI/ML approaches also raises security concerns, such as model vulnerabilities and poisoning, inference and adversarial attacks, and risks related to data privacy and integrity. Therefore, it is essential to identify the security risks related to AI models and mitigate those risks before deployment stages \cite{kaur2025applications}. 

Federated Learning (FL) is a distributed Artificial Intelligence (AI) technology, which is utilized in NextG networks to enhance security, particularly the trustworthiness of AI models. FL frameworks offer a promising paradigm for training machine learning models across distributed nodes, e.g., base stations, radar units, or user equipment, without centralizing raw data \cite{jain2025toward, 11389802,seo2026gc}. However, exchanging model updates over the network raises privacy and security concerns, i.e., an eavesdropper or a curious server may infer sensitive information from gradient or weight updates, and compromised keys can lead to insecure aggregation \cite{hou2025towards}. The authors in \cite{abasi20256g} propose an FL framework to ensure data integrity, privacy, and performance reliability for
millimeter-wave (mmWave) beam prediction systems in 6G. This study utilizes the adaptive noise augmentation and differential privacy principles to mitigate vulnerabilities in FL systems, providing a robust
defense against adversarial attacks, with an average of 17.45\% improvement across all scenarios. Another approach is the Quantum Key Distribution (QKD), utilizing the principles of quantum physics to improve the security in FL.  QKD enables information-theoretically secure key agreement, detecting eavesdropping via the quantum bit error rate (QBER) \cite{kumar2025brief}. Combined with cryptographic secure aggregation, QKD-derived keys can mask model updates so that the server obtains only the aggregated result while individual client contributions remain hidden. 

This paper proposes a framework that integrates a BB84-style QKD protocol abstraction with pairwise additive masking~\cite{bonawitz2017practical} for secure FL, applied to two representative NextG tasks: (1) channel estimation from pilot spectrograms and (2) radar spectrum sensing via spectrogram segmentation. Main contributions of this study are given as follows:
\begin{itemize}
\item Formalize a system model coupling QKD (BB84-style) with pairwise masking for secure FL, and derive the reconstruction correctness and QBER-based abort conditions.
\item Apply the framework to channel estimation (CNN on Rayleigh-fading pilots) and radar spectrum sensing (U-Net on spectrogram segmentation), demonstrating utility preservation under secure aggregation.
\item Evaluate the threat model when an eavesdropper is present, showing that QBER-based abort prevents aggregation with compromised keys and preserves security.
\item Characterize QBER sensitivity to depolarizing channel noise and report reconstruction error as well as leakage proxies (cosine and Pearson correlation) to validate the masking effectiveness.
\end{itemize}



\section{System Model}
\label{sec:system_model}

This section provides the details of the integrated system comprising (i) wireless channel estimation and radar spectrum sensing task models, (ii) a BB84-style quantum key distribution (QKD) protocol abstraction, (iii) pairwise additive masking for secure aggregation, and (iv) the federated learning framework. This study focuses on operating at the \emph{protocol level}; physical-layer measurement-device-independent (MDI) QKD modeling is out of scope.

\subsection{Channel Estimation Model}
\label{subsec:channel_model}

\begin{figure*}[htbp]
    \centering
    \begin{minipage}{0.35\linewidth}
        \centering
        \includegraphics[width=\linewidth]{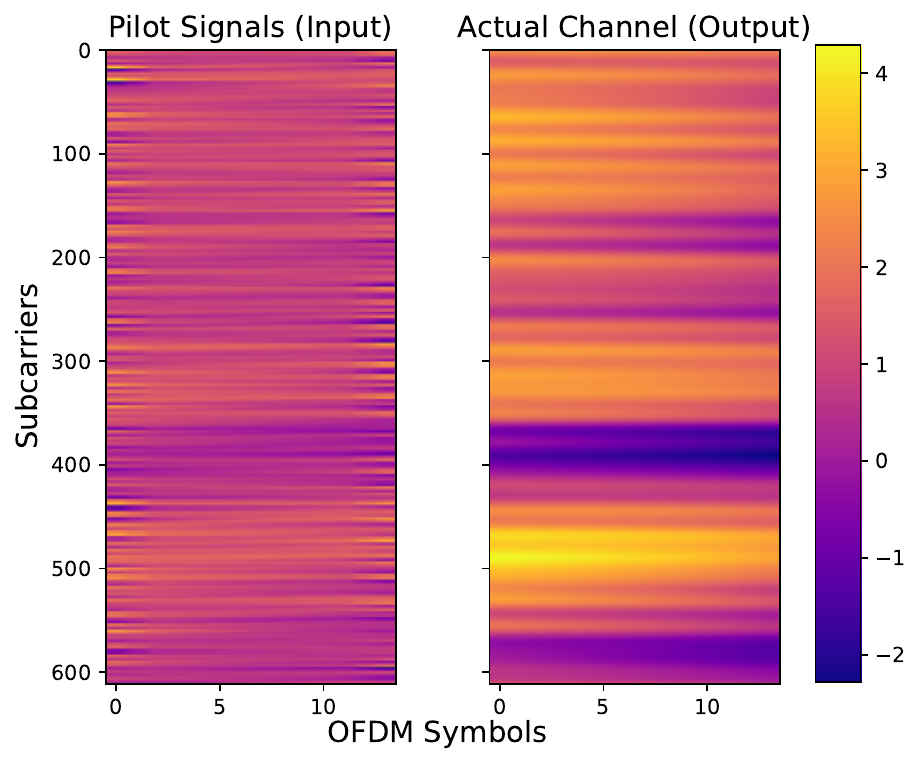}
    \caption{Channel Estimation Example data}
    \label{fig:placeholder_1}
    \end{minipage}\hfill
    \begin{minipage}{0.35\linewidth}
        \centering
        \includegraphics[width=\linewidth]{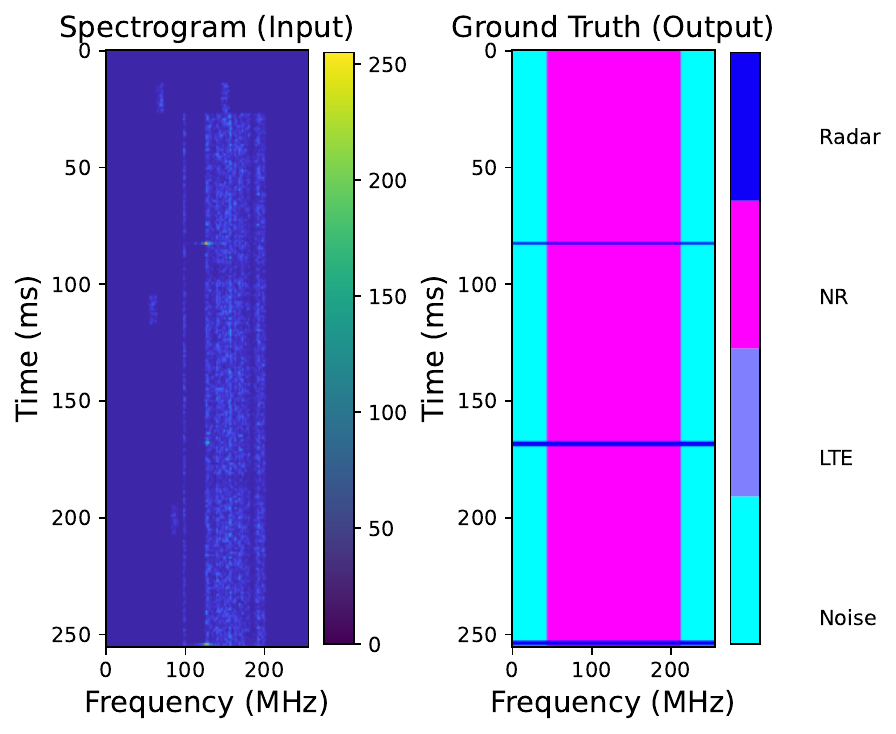}
    \caption{Radar Spectrum Example data}
    \label{fig:placeholder_redar}
    \end{minipage}
\end{figure*}

Figure \ref{fig:placeholder_1} illustrates the channel estimation example data. It is considered a single-user OFDM system with $N_{\mathrm{sub}}$ subcarriers and $N_{\mathrm{sym}}$ symbols per resource block. The received pilot signal at subcarrier $n$ and symbol $m$ is modeled as
\begin{equation}
\label{eq:channel_model}
Y_{n,m} = X_{n,m} \, H_{n,m} + Z_{n,m}, \quad n \in [N_{\mathrm{sub}}], \; m \in [N_{\mathrm{sym}}]
\end{equation}
where $X_{n,m} \in \mathbb{C}$ is the pilot symbol, $H_{n,m} \sim \mathcal{CN}(0, \sigma_h^2)$ is the complex channel gain (Rayleigh fading), and $Z_{n,m} \sim \mathcal{CN}(0, \sigma_z^2)$ is additive white Gaussian noise (AWGN). The signal-to-noise ratio is $\mathrm{SNR} = \sigma_h^2 / \sigma_z^2$.


The channel estimation task is to predict the channel magnitude $\lvert H_{n,m} \rvert$ from the noisy pilot observations. The input tensor is defined as
\begin{equation}
\mathbf{X} \in \mathbb{R}^{H \times W \times 1}, \quad H = 612, \; W = 14,
\end{equation}
representing a pilot spectrogram, and the target as
\begin{equation}
\mathbf{Y} = \lvert \mathbf{H} \rvert \in \mathbb{R}^{H \times W \times 1}.
\end{equation}

A convolutional neural network (CNN) $f_\theta: \mathbb{R}^{H \times W \times 1} \to \mathbb{R}^{H \times W \times 1}$ with parameters $\boldsymbol{\theta}$ maps pilots to channel magnitude estimates:
\begin{equation}
\hat{\mathbf{Y}} = f_\theta(\mathbf{X}), \quad \hat{Y}_{n,m} \approx \lvert H_{n,m} \rvert.
\end{equation}

The network architecture comprises three convolutional layers:
\begin{align}
\mathbf{h}_1 &= \sigma_{\mathrm{SELU}}\bigl(\mathrm{Conv}_{9\times9}^{(1 \to 48)}(\mathbf{X})\bigr), \\
\mathbf{h}_2 &= \sigma_{\mathrm{Softplus}}\bigl(\mathrm{Conv}_{5\times5}^{(48 \to 16)}(\mathbf{h}_1)\bigr), \\
\hat{\mathbf{Y}} &= \sigma_{\mathrm{SELU}}\bigl(\mathrm{Conv}_{5\times5}^{(16 \to 1)}(\mathbf{h}_2)\bigr).
\end{align}

Training minimizes the mean squared error (MSE) loss:
\begin{equation}
\mathcal{L}_{\mathrm{CE}}(\boldsymbol{\theta}) = \mathbb{E}_{(\mathbf{X},\mathbf{Y}) \sim \mathcal{D}} \Bigl[ \bigl\| f_\theta(\mathbf{X}) - \mathbf{Y} \bigr\|_2^2 \Bigr].
\end{equation}

The normalized mean squared error (NMSE) is used as the evaluation metric:
\begin{equation}
\label{eq:nmse}
\mathrm{NMSE} = \frac{\mathbb{E}\bigl[\| \hat{\mathbf{Y}} - \mathbf{Y} \|_2^2\bigr]}{\mathbb{E}\bigl[\| \mathbf{Y} \|_2^2\bigr] + \varepsilon}, \quad \varepsilon > 0.
\end{equation}

\subsection{Radar Spectrum Sensing Model}
\label{subsec:radar_model}

Figure \ref{fig:placeholder_redar} illustrates the radar example data. For radar spectrum sensing, the input is a received spectrogram $\mathbf{X}_{\mathrm{rad}} \in \mathbb{R}^{256 \times 256 \times 3}$ with three channels (e.g., time-frequency representations). The task is pixel-wise semantic segmentation into $C = 4$ classes: $\{\texttt{Noise}, \texttt{LTE}, \texttt{NR}, \texttt{Radar}\}$.

The target output is a label map $\mathbf{Y}_{\mathrm{rad}} \in \{0,1,2,3\}^{256 \times 256}$. A U-Net architecture $g_\phi: \mathbb{R}^{256 \times 256 \times 3} \to \mathbb{R}^{256 \times 256 \times C}$ with parameters $\boldsymbol{\phi}$ produces per-pixel class probabilities:
\begin{equation}
\hat{\mathbf{P}} = g_\phi(\mathbf{X}_{\mathrm{rad}}), \quad \hat{\mathbf{Y}}_{\mathrm{rad}} = \argmax_{c \in [C]} \hat{P}_{:,:,c}.
\end{equation}


The U-Net comprises an encoder-decoder structure with skip connections. Let $\mathcal{E}_\ell$ and $\mathcal{D}_\ell$ denote encoder and decoder blocks at level $\ell$. The encoder downsamples via
\begin{equation}
\mathbf{e}_\ell, \mathbf{p}_\ell = \mathcal{E}_\ell(\mathbf{p}_{\ell-1}), \quad \mathbf{p}_\ell = \mathrm{MaxPool}_{2\times2}(\mathbf{e}_\ell),
\end{equation}
with filter counts $(64, 128, 256, 512)$ and a bottleneck of 1024. The decoder upsamples and concatenates skip features:
\begin{equation}
\mathbf{d}_\ell = \mathcal{D}_\ell\bigl(\mathrm{UpSample}(\mathbf{d}_{\ell+1}), \mathbf{e}_\ell\bigr).
\end{equation}

Training uses categorical cross-entropy:
\begin{equation}
\mathcal{L}_{\mathrm{Radar}}(\boldsymbol{\phi}) = -\mathbb{E}_{(\mathbf{X},\mathbf{Y}) \sim \mathcal{D}_{\mathrm{rad}}} \sum_{i,j,c} \mathbb{1}[Y_{i,j} = c] \log \hat{P}_{i,j,c}.
\end{equation}

Model quality is evaluated by accuracy and mean intersection-over-union (mIoU):
\begin{equation}
\mathrm{mIoU} = \frac{1}{C} \sum_{c=0}^{C-1} \frac{|\hat{\mathcal{R}}_c \cap \mathcal{R}_c|}{|\hat{\mathcal{R}}_c \cup \mathcal{R}_c|},
\end{equation}
where $\mathcal{R}_c$ and $\hat{\mathcal{R}}_c$ are the ground-truth and predicted regions for class $c$.

\subsection{BB84-Style QKD Protocol Abstraction}
\label{subsec:qkd_model}

In this study, QKD is modeled at the protocol level as a black box that produces a shared secret key between two parties (e.g., each client and the server). The BB84 protocol uses two mutually unbiased bases: rectilinear $\{\ket{0}, \ket{1}\}$ and diagonal $\{\ket{+}, \ket{-}\}$ where $\ket{\pm} = (\ket{0} \pm \ket{1})/\sqrt{2}$.

\textbf{State preparation.} Alice encodes bit $b \in \{0,1\}$ in basis $a \in \{0,1\}$:
\begin{equation}
\rho_{b,a} = \begin{cases}
 \ket{0}\bra{0} \text{ or } \ket{1}\bra{1} & a = 0 \text{ (rectilinear)}, \\
 \ket{+}\bra{+} \text{ or } \ket{-}\bra{-} & a = 1 \text{ (diagonal)}.
\end{cases}
\end{equation}

\textbf{Measurement.} Bob measures in randomly chosen basis $\beta \in \{0,1\}$. The outcome is a bit $\tilde{b}$; when $\beta = a$, $\tilde{b} = b$ in the ideal noiseless case.

\textbf{Sifting.} Only indices $i$ where $a_i = \beta_i$ are retained. The sifted key length is $L_{\mathrm{sift}} \approx \ell/2$ for raw key length $\ell$.

\textbf{Quantum bit error rate (QBER).} Define
\begin{equation}
\label{eq:qber}
\mathrm{QBER} = \frac{1}{L_{\mathrm{sift}}} \sum_{i \in \mathcal{I}_{\mathrm{sift}}} \mathbb{1}[b_i \neq \tilde{b}_i].
\end{equation}

The information-theoretic security of BB84 has been formally established under realistic assumptions \cite{PhysRevLett.85.441}, linking QBER thresholds to provable key secrecy. If $\mathrm{QBER} \geq \tau_{\mathrm{QBER}}$ (e.g., $\tau_{\mathrm{QBER}} = 0.11$), the round is aborted due to possible eavesdropping .

\textbf{Privacy amplification.} A hash-based extractor $\mathsf{PA}: \{0,1\}^{L_{\mathrm{sift}}} \to \{0,1\}^{L_{\mathrm{final}}}$ with $L_{\mathrm{final}} = \max\bigl(256, \lfloor \rho_{\mathrm{PA}} \cdot L_{\mathrm{sift}} \rfloor\bigr)$, $\rho_{\mathrm{PA}} \approx 0.8$, produces the final key $\mathbf{k} \in \{0,1\}^{L_{\mathrm{final}}}$.

\subsection{QKD-Seeded Pairwise Masking for Secure Aggregation}

In the implemented framework, a single QKD-derived secret seed is established per federated learning round. Let
\begin{equation}
k^{(r)} \in \{0,1\}^{L}
\end{equation}
denote the shared secret key generated via the BB84 protocol abstraction at round $r$, where $L$ is the final key length after privacy amplification.

Instead of performing independent QKD sessions for each client pair, we deterministically derive pairwise masking keys from the round seed using a key-derivation function (KDF):
\begin{equation}
k_{ij}^{(r)} = \mathrm{KDF}\big(k^{(r)}, \min(i,j), \max(i,j)\big),
\end{equation}
for all client pairs $(i,j)$ with $i \neq j$. By construction,
\begin{equation}
k_{ij}^{(r)} = k_{ji}^{(r)},
\end{equation}
ensuring symmetric masking between client pairs. In practice, the KDF is implemented using a cryptographic hash-based expansion (e.g., SHA-256 in counter mode) to produce sufficient key material for mask generation.

From each derived key $k_{ij}^{(r)}$, a masking tensor $m_{ij}$ is generated by mapping key bits deterministically to $\pm \gamma$, where $\gamma > 0$ is a small masking scale factor. The mapping is applied cyclically over the flattened parameter tensor to match the dimensionality of $\theta$.

Client $i$ constructs its masked update as
\begin{equation}
\tilde{\theta}^{(i)} =
\theta^{(i)} +
\sum_{j>i} m_{ij}
-
\sum_{j<i} m_{ji}.
\label{eq:masked_params}
\end{equation}

Because $m_{ij} = m_{ji}$, the masks cancel during aggregation:
\begin{equation}
\sum_{i=1}^{K} \tilde{\theta}^{(i)}
=
\sum_{i=1}^{K} \theta^{(i)} +
\sum_{i<j} (m_{ij} - m_{ij})
=
\sum_{i=1}^{K} \theta^{(i)}.
\end{equation}

Thus, the server obtains the correct global average without observing individual unmasked updates. Security relies on the secrecy of the QKD-derived round seed $k^{(r)}$ and the collision resistance of the KDF, while aggregation correctness follows from deterministic mask symmetry and cancellation.

\subsection{Federated Learning Framework}
\label{subsec:fl_framework}

The global model parameters $\boldsymbol{\theta}^{\mathrm{global}}$ are updated over $R$ rounds. In round $r$:

\begin{enumerate}
\item \textbf{Key establishment.} Run QKD once per federated learning round to obtain a shared secret seed 
$k^{(r)}$ and estimate the QBER. 
If $\mathrm{QBER} \ge \tau_{\mathrm{QBER}}$, abort the round. 
Otherwise, derive pairwise masking keys
\[
k^{(r)}_{ij} = \mathrm{KDF}\big(k^{(r)}, \min(i,j), \max(i,j)\big)
\]
for all client pairs $(i,j)$.
\item \textbf{Broadcast.} Server sends $\boldsymbol{\theta}^{\mathrm{global}}$ to all clients.
\item \textbf{Local training.} Each client $k$ trains on local data $\mathcal{D}_k$ for $E$ epochs:
\begin{equation}
\boldsymbol{\theta}^{(k)}_r \leftarrow \mathrm{LocalTrain}\bigl(\boldsymbol{\theta}^{\mathrm{global}}, \mathcal{D}_k, E\bigr).
\end{equation}
\item \textbf{Masking \& upload.} Client $k$ computes $\tilde{\boldsymbol{\theta}}^{(k)}_r$ via Equation~\eqref{eq:masked_params} and sends it to the server.
\item \textbf{Aggregation.} Server computes $\bar{\boldsymbol{\theta}}_r = \frac{1}{K}\sum_{k=1}^{K} \tilde{\boldsymbol{\theta}}^{(k)}_r$ and sets $\boldsymbol{\theta}^{\mathrm{global}} \leftarrow \bar{\boldsymbol{\theta}}_r$.
\end{enumerate}

Local training minimizes the empirical risk
\begin{equation}
\min_{\boldsymbol{\theta}} \; \frac{1}{|\mathcal{D}_k|} \sum_{(\mathbf{x}, \mathbf{y}) \in \mathcal{D}_k} \mathcal{L}\bigl(f_\theta(\mathbf{x}), \mathbf{y}\bigr),
\end{equation}
using stochastic gradient descent (e.g., Adam) with learning rate $\eta$.

\subsection{Integrated System Architecture}
\label{subsec:architecture}

\begin{figure*}[!htbp]
\centering
\includegraphics[width=0.75\linewidth]{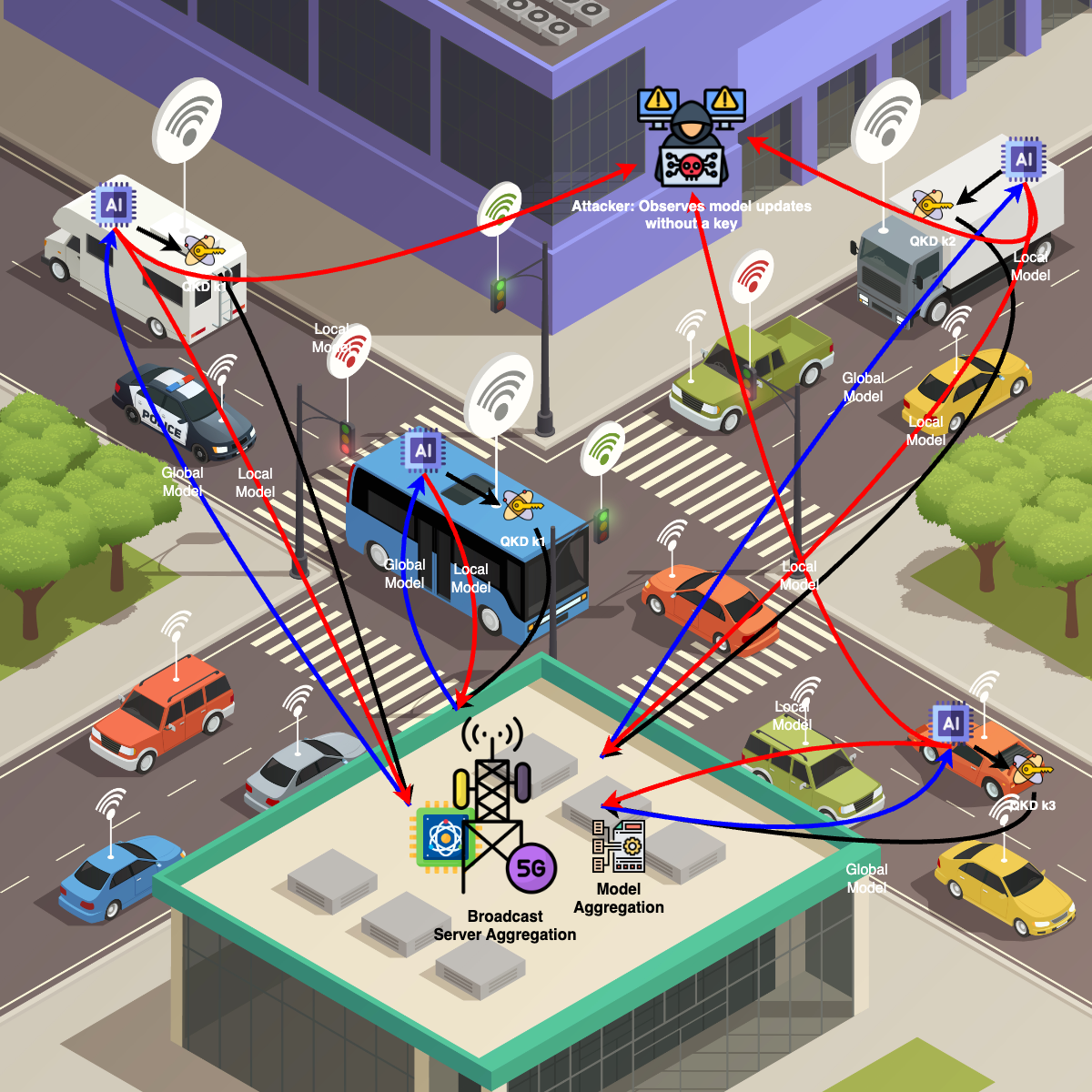}
\caption{QKD-secured federated learning system architecture. Clients train locally, mask updates with pairwise keys, and upload; server aggregates without seeing plain parameters.}
\label{fig:system_architecture}
\end{figure*}

Figure~\ref{fig:system_architecture} illustrates the end-to-end flow. Multiple clients hold non-IID local datasets for channel estimation or radar sensing. Each client trains a local model (CNN or U-Net), obtains pairwise QKD keys with the server (or with other clients), masks the model update, and uploads only the masked parameters. The server aggregates and broadcasts the new global model. An eavesdropper observing the masked updates cannot recover individual model parameters without the secret keys.


\begin{figure}[!htbp]
\centering
\includegraphics[width=1.0\columnwidth]{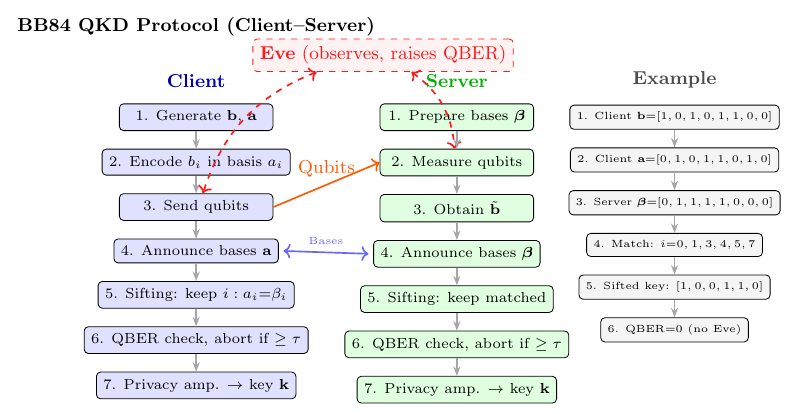}
\caption{BB84 QKD protocol flow: state preparation, measurement, sifting, QBER check, and privacy amplification.}
\label{fig:bb84_flow}
\end{figure}

\begin{figure}[!htbp]
\centering
\includegraphics[width=\columnwidth]{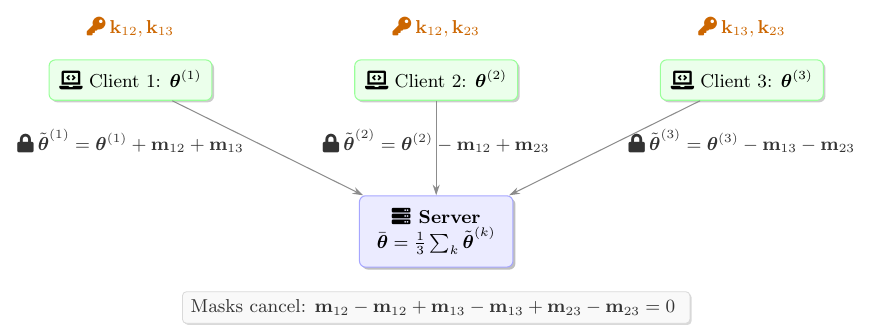}
\caption{Pairwise additive masking: each client adds masks for $j>i$ and subtracts for $j<i$; masks cancel at aggregation.}
\label{fig:pairwise_masking}
\end{figure}



\begin{figure*}[htbp]
    \centering
    \begin{minipage}{0.30\linewidth}
        \centering
        \includegraphics[width=\linewidth]{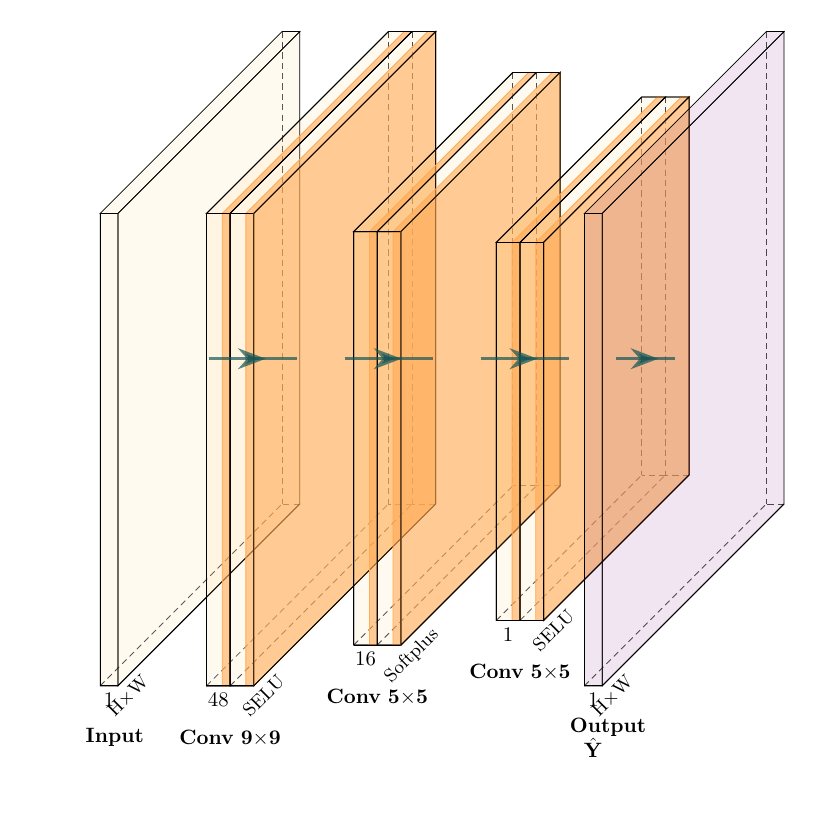}
        \caption{Channel estimation CNN: three convolutional layers with SELU and Softplus activations.}
        \label{fig:channel_cnn}
    \end{minipage}\hfill
    \begin{minipage}{0.45\linewidth}
        \centering
        \includegraphics[width=\linewidth]{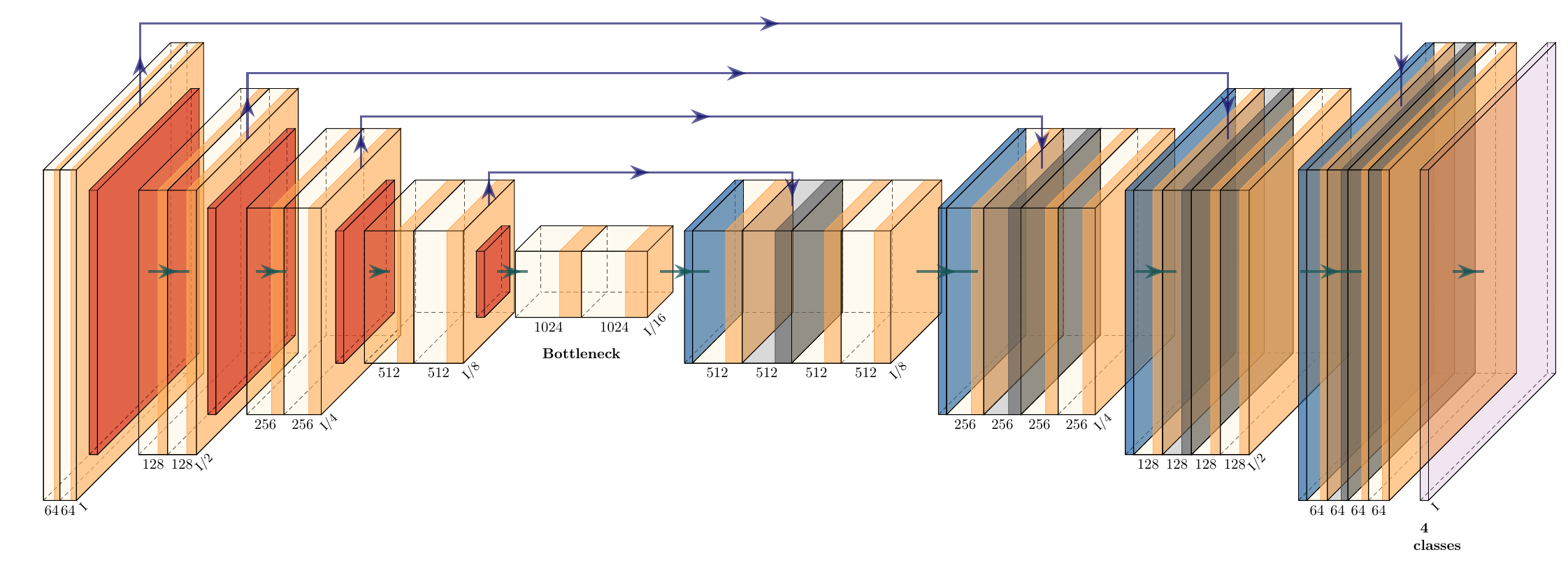}
        \caption{U-Net architecture for radar spectrum sensing: encoder-decoder with skip connections.}
        \label{fig:unet}
    \end{minipage}
\end{figure*}

\begin{figure}[!htbp]
\centering
\includegraphics[width=1.0\columnwidth]{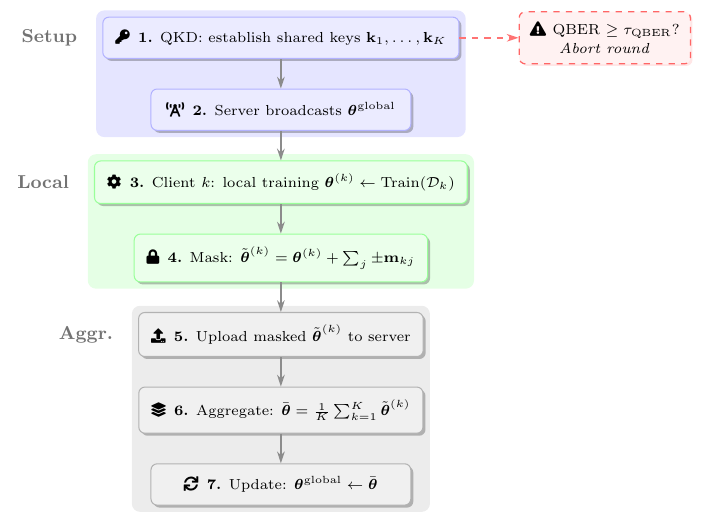}
\caption{One federated learning round with QKD-secured aggregation and QBER-based abort.}
\label{fig:fl_round}
\end{figure}

\subsection{Threat Model}

We consider a semi-honest server and passive eavesdroppers.
The server follows the federated learning protocol but may
attempt to infer client data from received model updates.
Eavesdroppers can observe all transmitted masked parameters
but do not possess secret key material.

In the implemented framework, a single QKD-derived secret
seed $k^{(r)}$ is established once per federated learning round.
This seed is shared among participating clients but is not
accessible to the aggregation server. Pairwise masking keys
are deterministically derived from the round seed using a
cryptographic key-derivation function (KDF):

\begin{equation}
k_{ij}^{(r)} = \mathrm{KDF}\!\left(k^{(r)}, \min(i,j), \max(i,j)\right),
\end{equation}

where $k_{ij}^{(r)} = k_{ji}^{(r)}$.
The KDF is implemented using a hash-based expansion
(e.g., SHA-256 counter mode) to generate sufficient key bits
for mask construction.

Security of aggregation relies on the secrecy of the QKD-derived
round seed $k^{(r)}$ and the collision resistance of the KDF.
Since the server does not possess $k^{(r)}$, it cannot reconstruct
individual pairwise masking keys and therefore cannot recover
plain client updates. Mask cancellation ensures that only the
aggregated model is revealed.

If the estimated QBER exceeds the threshold $\tau_{\mathrm{QBER}}$,
the round is aborted and no key material is used.


\subsection{Algorithms}
\label{subsec:algorithms}

Algorithm~\ref{alg:bb84} outlines the BB84 key generation. Algorithm~\ref{alg:pairwise_mask} details pairwise masking, and Algorithm~\ref{alg:fl_round} describes one federated learning round with QKD-secured aggregation.

\begin{algorithm}[!htbp]
\small
\caption{BB84 Key Generation (Protocol Abstraction)}
\label{alg:bb84}
\begin{algorithmic}[1]
\REQUIRE Raw key length $\ell$, privacy amplification ratio $\rho_{\mathrm{PA}}$, optional: channel noise $\eta$
\ENSURE Shared key $\mathbf{k}$, sifted length $L_{\mathrm{sift}}$, final length $L_{\mathrm{final}}$, QBER
\STATE Alice samples $\mathbf{b}_A \in \{0,1\}^\ell$, $\mathbf{a}_A \in \{0,1\}^\ell$ (bits and bases)
\STATE Bob samples $\boldsymbol{\beta}_B \in \{0,1\}^\ell$ (measurement bases)
\STATE Bob measures and obtains $\tilde{\mathbf{b}}$
\STATE $\mathcal{I}_{\mathrm{sift}} \gets \{ i : a_{A,i} = \beta_{B,i} \}$
\STATE $\mathbf{b}_{\mathrm{sift}} \gets (\tilde{b}_i)_{i \in \mathcal{I}_{\mathrm{sift}}}$, $L_{\mathrm{sift}} \gets |\mathcal{I}_{\mathrm{sift}}|$
\STATE $\mathrm{QBER} \gets \frac{1}{L_{\mathrm{sift}}} \sum_{i \in \mathcal{I}_{\mathrm{sift}}} \mathbb{1}[b_{A,i} \neq \tilde{b}_i]$
\STATE $L_{\mathrm{final}} \gets \max(256, \lfloor \rho_{\mathrm{PA}} \cdot L_{\mathrm{sift}} \rfloor)$
\STATE $\mathbf{k} \gets \mathsf{PA}(\mathbf{b}_{\mathrm{sift}}, L_{\mathrm{final}})$
\RETURN $\mathbf{k}$, $L_{\mathrm{sift}}$, $L_{\mathrm{final}}$, QBER
\end{algorithmic}
\end{algorithm}

\begin{algorithm}[!htbp]
\small
\caption{Pairwise Masking}
\label{alg:pairwise_mask}
\begin{algorithmic}[1]
\REQUIRE Local state dict $\boldsymbol{\theta}^{(i)}$, client index $i$, round $r$, number of clients $K$, base seed $s$, key bit length $L$, scale $\gamma$
\ENSURE Masked state dict $\tilde{\boldsymbol{\theta}}^{(i)}$
\STATE $\tilde{\boldsymbol{\theta}}^{(i)} \gets \boldsymbol{\theta}^{(i)}$
\FOR{each parameter name $n$ in $\boldsymbol{\theta}^{(i)}$}
    \FOR{$j = 0$ to $K-1$}
        \IF{$j = i$}
            \STATE \textbf{continue}
        \ENDIF
        \STATE $\mathbf{k}_{ij} \gets \mathsf{DeriveKey}(i, j, r, s, L)$
        \STATE $\mathbf{m}_{ij} \gets \mathsf{BitsToMask}(\mathbf{k}_{ij}, \mathrm{shape}(\boldsymbol{\theta}^{(i)}_n), \gamma)$
        \IF{$i < j$}
            \STATE $\tilde{\boldsymbol{\theta}}^{(i)}_n \gets \tilde{\boldsymbol{\theta}}^{(i)}_n + \mathbf{m}_{ij}$
        \ELSE
            \STATE $\tilde{\boldsymbol{\theta}}^{(i)}_n \gets \tilde{\boldsymbol{\theta}}^{(i)}_n - \mathbf{m}_{ij}$
        \ENDIF
    \ENDFOR
\ENDFOR
\RETURN $\tilde{\boldsymbol{\theta}}^{(i)}$
\end{algorithmic}
\end{algorithm}

\begin{algorithm}[!htbp]
\small
\caption{QKD-Seeded Secure Aggregation Federated Learning Round}
\label{alg:fl_round}
\begin{algorithmic}[1]
\REQUIRE Global model $\boldsymbol{\theta}^{\mathrm{global}}$, clients $\{1,\ldots,K\}$, local data $\{\mathcal{D}_1,\ldots,\mathcal{D}_K\}$, round index $r$
\ENSURE Updated global model $\boldsymbol{\theta}^{\mathrm{global}}$, metrics
\STATE Run QKD once for the round to obtain a shared seed key $\mathbf{k}^{(r)}$ and QBER
\IF{QBER $\geq \tau_{\mathrm{QBER}}$}
    \STATE \textbf{abort} round; return status ABORTED
\ENDIF
\FOR{$k = 1$ to $K$}
    \STATE $\boldsymbol{\theta}^{(k)} \gets \mathrm{LocalTrain}(\boldsymbol{\theta}^{\mathrm{global}}, \mathcal{D}_k, E)$
    \STATE \hspace{1em}\textit{// Inside $\mathsf{ApplyPairwiseMasks}$: derive pairwise keys } $\mathbf{k}_{kj}^{(r)}=\mathrm{KDF}(\mathbf{k}^{(r)},k,j)$ \textit{ and map to } $\pm\gamma$ \textit{ masks}
    \STATE $\tilde{\boldsymbol{\theta}}^{(k)} \gets \mathsf{ApplyPairwiseMasks}\!\left(\boldsymbol{\theta}^{(k)}, k, r, K, \mathbf{k}^{(r)}, L, \gamma\right)$
    \STATE Client $k$ sends $\tilde{\boldsymbol{\theta}}^{(k)}$ to server
\ENDFOR
\STATE $\bar{\boldsymbol{\theta}} \gets \frac{1}{K} \sum_{k=1}^{K} \tilde{\boldsymbol{\theta}}^{(k)}$
\STATE $\boldsymbol{\theta}^{\mathrm{global}} \gets \bar{\boldsymbol{\theta}}$
\STATE Evaluate $\boldsymbol{\theta}^{\mathrm{global}}$ on validation set; compute NMSE / accuracy / mIoU
\RETURN $\boldsymbol{\theta}^{\mathrm{global}}$, metrics
\end{algorithmic}
\end{algorithm}


\section{Experimental Results}
\label{sec:experiments}

The QKD-secured federated learning framework is evaluated in terms of two tasks: (i) channel estimation on OFDM pilot data, and (ii) radar spectrum sensing on spectrogram segmentation. The same datasets are used as in our prior work \cite{catak2024spectrumm, catak2023crypto}. Three experiment types are conducted: \textbf{Experiment A} compares baseline FL (plain aggregation) with secure FL (masked aggregation); \textbf{Experiment B} assesses the threat model when an eavesdropper is present; \textbf{Experiment C} sweeps depolarizing channel noise to characterize QBER sensitivity and abort behavior.

\subsection{Dataset Description}
\label{subsec:datasets}

\subsubsection{Channel Estimation Dataset}
The channel estimation dataset is generated from the ``Deep Learning Data Synthesis for 5G Channel Estimation'' reference model in the MATLAB 5G Toolbox \footnote{\url{https://www.mathworks.com/products/5g.html}}. Each sample consists of transmit and receive signals with $612$ subcarriers and $14$ OFDM symbols (single antenna), yielding $8568$ complex-valued data points per instance. The complex channel response is converted to real-valued inputs (magnitude or real/imaginary) of size $612 \times 14 \times 1$ for the CNN. Table~\ref{tab:ce_dataset} summarizes the dataset.

\begin{table}[t]
\centering
\caption{Channel Estimation Dataset}
\label{tab:ce_dataset}
\begin{tabular}{lc}
\toprule
Parameter & Value \\
\midrule
Source & MATLAB 5G Toolbox (synthesis) \\
Image size & $612 \times 14$ \\
Training samples & 1000 \\
Validation samples & 500 \\
\bottomrule
\end{tabular}
\end{table}

\subsubsection{Radar Spectrum Sensing Dataset}
The radar spectrum sensing dataset is synthesized using deep learning to enable semantic segmentation in environments where radar and wireless communication systems (LTE, 5G NR) coexist. The dataset is generated with MATLAB toolboxes: Phased Array System Toolbox, 5G Toolbox, LTE Toolbox, Computer Vision Toolbox, and Deep Learning Toolbox. The objective is to train a model to differentiate radar and wireless communication signals within a shared spectral domain. Table~\ref{tab:radar_params} gives the radar signal parameters; Table~\ref{tab:radar_scenario} describes the scenario setup for channel modeling.

\begin{table}[t]
\centering
\caption{Radar Signal Parameters}
\label{tab:radar_params}
\begin{tabular}{lc}
\toprule
Parameter & Value \\
\midrule
Center frequency ($f_c$) & 2.8 GHz \\
Sampling frequency ($f_s$) & 61.44 MHz \\
Pulse repetition frequency & Variable \\
Pulse width & 1 $\mu$s \\
Transmit power & 25 kW \\
Antenna gain & 32.8 dB \\
\bottomrule
\end{tabular}
\end{table}

\begin{table}[t]
\centering
\caption{Scenario Setup for Channel Modeling (Radar Dataset)}
\label{tab:radar_scenario}
\begin{tabular}{ll}
\toprule
Property & Value \\
\midrule
Transmitter positions & Random within 2 km $\times$ 2 km \\
Number of scatterers & 30 \\
Direct path & Disabled \\
Atmospheric conditions & Simulated \\
\bottomrule
\end{tabular}
\end{table}

Wireless communication signals include 5G NR and LTE; transmitter locations vary randomly across frames. The spectrograms are of size $256 \times 256 \times 3$ with pixel-wise labels in $\{0,1,2,3\}$ corresponding to Noise, LTE, NR, and Radar. Table~\ref{tab:rad_dataset} summarizes the dataset used in this study.

\begin{table}[t]
\centering
\caption{Radar Spectrum Sensing Dataset}
\label{tab:rad_dataset}
\begin{tabular}{lc}
\toprule
Parameter & Value \\
\midrule
Spectrogram size & $256 \times 256 \times 3$ \\
Classes & Noise, LTE, NR, Radar \\
Total samples & 200 \\
Train/validation split & 80\%/20\% \\
\bottomrule
\end{tabular}
\end{table}

\subsection{Experimental Setup}
\label{subsec:setup}

\textbf{Channel estimation:} The CNN has $48 \to 16 \to 1$ filters (Conv2D layers); local training uses Adam with learning rate $\eta = 10^{-3}$, $E = 3$ epochs per round, and batch size 16. We study scalability over the number of FL clients $K \in \{3,10,20\}$ and run $R = 5$ FL rounds for each configuration. For each pair $(K,r)$ we consider three aggregation modes: \emph{plain} (no masking), \emph{Classical-SA} (pairwise masking with PRG-derived keys), and \emph{QKD-SA} (pairwise masking with BB84-derived keys). The channel-estimation loss is normalized MSE
\begin{equation}
  \mathrm{NMSE}
  = \frac{\|\mathbf{h} - \hat{\mathbf{h}}\|_2^2}{\|\mathbf{h}\|_2^2},
\end{equation}
where $\mathbf{h}$ is the true channel and $\hat{\mathbf{h}}$ is the CNN estimate.

\textbf{Radar sensing:} The U-Net has $\sim$31M parameters; local training uses Adam with $\eta = 10^{-4}$, $E = 3$ epochs, and batch size 4. As in channel estimation, we run $R = 5$ FL rounds for each $K \in \{3,10,20\}$ and for the three aggregation modes. The primary utility metrics are pixel accuracy
\begin{equation}
  \mathrm{Acc}
  = \frac{1}{N}\sum_{i=1}^{N} \mathbf{1}\{\hat{y}_i = y_i\},
\end{equation}
and mean intersection-over-union (mIoU)
\begin{equation}
  \mathrm{mIoU}
  = \frac{1}{|\mathcal{C}|}\sum_{c \in \mathcal{C}}
  \frac{|\hat{\mathcal{Y}}_c \cap \mathcal{Y}_c|}{|\hat{\mathcal{Y}}_c \cup \mathcal{Y}_c|},
\end{equation}
where $\mathcal{C}$ is the set of semantic classes, $\hat{\mathcal{Y}}_c$ and $\mathcal{Y}_c$ denote predicted and true pixel sets for class $c$, and $N$ is the total number of pixels.

\textbf{QKD \& masking:} Raw key length $\ell = 2000$ bits; privacy amplification ratio $\rho_{\mathrm{PA}} \approx 0.8$. We adopt a QBER threshold $\tau_{\mathrm{QBER}} = 0.08$ for both tasks; any round with estimated QBER $\geq \tau_{\mathrm{QBER}}$ is aborted. The pairwise mask scale is set to $\gamma = 10^{-3}$ so that masks are large enough to obfuscate individual updates while remaining numerically stable for aggregation.

\subsection{Experiment A: Utility and Scalability}
\label{subsec:exp_a}

Experiment A compares the three aggregation modes for different numbers of clients and quantifies the impact on utility and communication. For each $K \in \{3,10,20\}$ and mode $m \in \{\text{plain},\text{Classical-SA},\text{QKD-SA}\}$, we record the final-round NMSE (channel estimation), accuracy and mIoU (radar sensing), as well as byte-level communication statistics.

\begin{table}[t]
\centering
\caption{Channel Estimation---Experiment A: Utility and Communication vs.\ Number of Clients}
\label{tab:ce_utility}
\begin{tabular}{lccc}
\toprule
$(K,\text{mode})$ & Final NMSE & Downlink [bytes] & Uplink [bytes] \\
\midrule
$(3,\text{plain})$        & $5.03\times 10^{-2}$ & $9.4\times 10^{4}$  & $2.83\times 10^{5}$ \\
$(3,\text{Classical-SA})$ & $6.25\times 10^{-2}$ & $9.4\times 10^{4}$  & $2.83\times 10^{5}$ \\
$(3,\text{QKD-SA})$       & $5.10\times 10^{-2}$ & $9.4\times 10^{4}$  & $2.83\times 10^{5}$ \\
$(10,\text{plain})$       & $6.56\times 10^{-2}$ & $9.4\times 10^{4}$  & $9.42\times 10^{5}$ \\
$(10,\text{QKD-SA})$      & $7.13\times 10^{-2}$ & $9.4\times 10^{4}$  & $9.42\times 10^{5}$ \\
$(20,\text{plain})$       & $8.60\times 10^{-2}$ & $9.4\times 10^{4}$  & $1.88\times 10^{6}$ \\
$(20,\text{QKD-SA})$      & $9.05\times 10^{-2}$ & $9.4\times 10^{4}$  & $1.88\times 10^{6}$ \\
\bottomrule
\end{tabular}
\end{table}

From Table~\ref{tab:ce_utility}, we observe that for $K=3$ the QKD-SA mode attains nearly identical NMSE to the plain baseline (about $5.1\times 10^{-2}$ vs.\ $5.0\times 10^{-2}$). As $K$ grows to $10$ and $20$, NMSE naturally increases due to the more heterogeneous client population; nevertheless, QKD-SA closely tracks the plain FL performance at each $K$. This confirms that pairwise masking does not introduce aggregation bias in practice: the masks cancel correctly at the server, and the global model quality is preserved.

Table~\ref{tab:rad_utility} reports radar sensing results. For $K=3$, QKD-SA attains the highest utility among the three modes, with pixel accuracy of approximately $89.0\%$ and mIoU of $0.68$, slightly outperforming the plain FL baseline ($\approx 88.1\%$ accuracy and $0.65$ mIoU). For larger numbers of clients ($K=10,20$), performance degrades due to increased non-iid heterogeneity; however, QKD-SA remains close to or better than the plain mode in most cases, particularly in mIoU.

\begin{table}[t]
\centering
\caption{Radar Sensing---Experiment A: Utility and Communication vs.\ Number of Clients}
\label{tab:rad_utility}
\begin{tabular}{lcccc}
\toprule
$(K,\text{mode})$ & Pixel Acc. & mIoU & Downlink [bytes] & Uplink [bytes] \\
\midrule
$(3,\text{plain})$        & $88.1\%$ & $0.65$ & $1.24\times 10^{8}$ & $3.73\times 10^{8}$ \\
$(3,\text{QKD-SA})$       & $89.0\%$ & $0.68$ & $1.24\times 10^{8}$ & $3.73\times 10^{8}$ \\
$(10,\text{plain})$       & $66.9\%$ & $0.41$ & $1.24\times 10^{8}$ & $1.24\times 10^{9}$ \\
$(10,\text{QKD-SA})$      & $82.9\%$ & $0.56$ & $1.24\times 10^{8}$ & $1.24\times 10^{9}$ \\
$(20,\text{plain})$       & $77.5\%$ & $0.47$ & $1.24\times 10^{8}$ & $2.48\times 10^{9}$ \\
$(20,\text{QKD-SA})$      & $73.7\%$ & $0.44$ & $1.24\times 10^{8}$ & $2.48\times 10^{9}$ \\
\bottomrule
\end{tabular}
\end{table}

\begin{figure*}[htbp]
    \centering
    \begin{minipage}{0.40\linewidth}
        \centering
        \includegraphics[width=\linewidth]{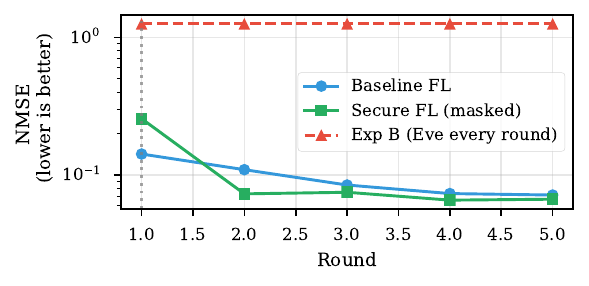}
        \caption{Channel estimation: NMSE versus FL round for baseline and secure (masked) modes.}
        \label{fig:ce_learning}
    \end{minipage}\hfill
    \begin{minipage}{0.40\linewidth}
        \centering
        \includegraphics[width=\linewidth]{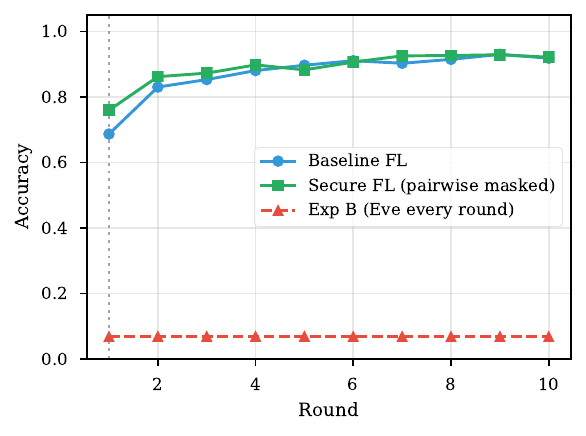}
        \caption{Radar sensing: accuracy and mIoU versus FL round for baseline and secure modes.}
        \label{fig:rad_learning}
    \end{minipage}
\end{figure*}

Figure~\ref{fig:ce_learning} shows NMSE evolution over rounds for channel estimation; Figure~\ref{fig:rad_learning} shows accuracy and mIoU for radar sensing. Across all $K$, the secure (QKD-SA) curves closely follow, and in some cases slightly outperform, the plain FL baseline, demonstrating that QKD-secured aggregation preserves model quality even in multi-client 6G scenarios.

\subsection{Experiment B: Threat Model Outcomes}
\label{subsec:exp_b}

Experiment B injects an eavesdropper (Eve) in every round, modeling an intercept-resend attack. When Eve is present, QBER rises and keys may diverge; the protocol aborts rounds when $\mathrm{QBER} \geq \tau_{\mathrm{QBER}}$.

Table~\ref{tab:threat_ce} shows threat model outcomes for channel estimation. With Eve present in all rounds, every round is aborted ($n_{\mathrm{aborted}} = 5$); no secure aggregation occurs. Baseline and secure modes (without Eve) complete all rounds successfully.

\begin{table}[t]
\centering
\caption{Channel Estimation---Experiment B: Threat Model}
\label{tab:threat_ce}
\begin{tabular}{lcccc}
\toprule
Mode & Rounds & Secure & Aborted & Recovered \\
\midrule
Baseline & 5 & 5 & 0 & 0 \\
Secure   & 5 & 5 & 0 & 0 \\
Eve (all rounds) & 5 & 0 & 5 & 0 \\
\bottomrule
\end{tabular}
\end{table}

Table~\ref{tab:threat_rad} summarizes the radar threat model. When rounds abort, the model retains only the prior global state; mean accuracy and mIoU under ABORTED reflect the untrained (or partially trained) model. The QBER-based abort successfully prevents use of compromised keys.

\begin{table}[t]
\centering
\caption{Radar Sensing---Experiment B: Threat Model Outcomes}
\label{tab:threat_rad}
\begin{tabular}{lcccc}
\toprule
Outcome & Mean QBER & Mean Acc. & Mean mIoU & Notes \\
\midrule
SECURE  & --- & --- & --- & Aggregation  \\
  &  &  &  &  proceeds \\
ABORTED & 24.6\% & 6.96\% & 0.0174 & All rounds  \\
 &  &  &  & aborted \\
\bottomrule
\end{tabular}
\end{table}

Table~\ref{tab:key_abort} details per-round abort status when Eve is present. Each round exceeds the QBER threshold and is aborted; model accuracy remains at the initial (random) level, demonstrating that the protocol correctly refuses to aggregate under attack.

\begin{table}[t]
\centering
\caption{Radar Sensing---Key Abort Effectiveness (Eve Present)}
\label{tab:key_abort}
\begin{tabular}{cccc}
\toprule
Round & Status & QBER & Acc. Retained \\
\midrule
1--10 & ABORTED & 23.0--26.4\% & 6.96\% \\
\bottomrule
\end{tabular}
\end{table}

\subsection{Experiment C: QBER Sensitivity}
\label{subsec:exp_c}

\begin{figure*}[htbp]
    \centering
    \begin{minipage}{0.40\linewidth}
        \centering
        \includegraphics[width=\linewidth]{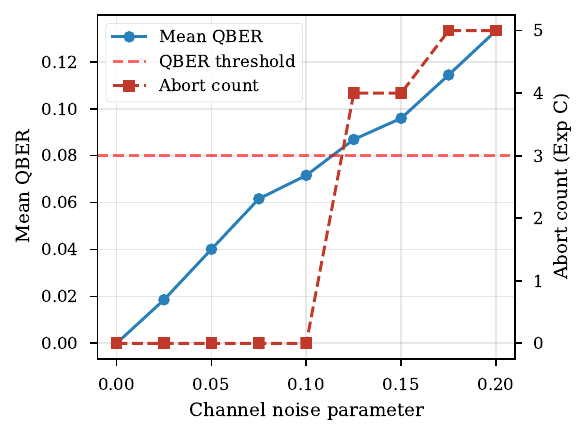}
        \caption{Channel estimation: QBER sensitivity to depolarizing noise $\eta$; abort when QBER $\geq 0.08$.}
        \label{fig:ce_qber}
    \end{minipage}\hfill
    \begin{minipage}{0.40\linewidth}
        \centering
        \includegraphics[width=\linewidth]{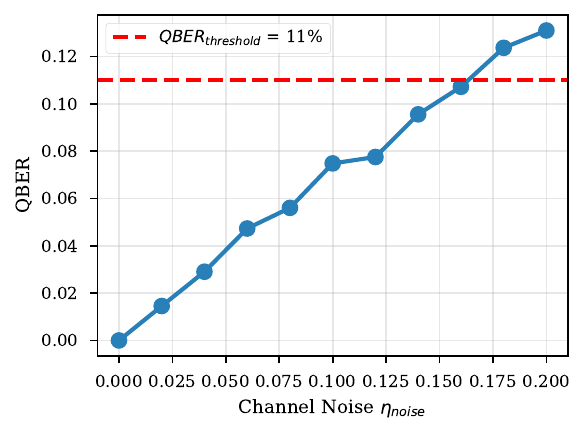}
        \caption{Radar sensing: QBER sensitivity to depolarizing noise $\eta$; abort when QBER $\geq 0.11$.}
        \label{fig:rad_qber}
    \end{minipage}
\end{figure*}

Experiment C sweeps depolarizing channel noise $\eta \in [0, 0.20]$ to characterize how QBER and abort rate scale with quantum channel degradation. Figures~\ref{fig:ce_qber} and~\ref{fig:rad_qber} plot mean QBER and abort count versus noise. As $\eta$ increases, QBER rises monotonically; when $\mathrm{QBER} \geq \tau_{\mathrm{QBER}}$, rounds abort. The abort threshold provides a clear security boundary: keys above the threshold are discarded, preventing aggregation with compromised keys.

\subsection{Reconstruction Correctness and Leakage Proxies}
\label{subsec:recon_leakage}

\begin{figure*}[htbp]
    \centering
    \begin{minipage}{0.40\linewidth}
        \centering
        \includegraphics[width=\linewidth]{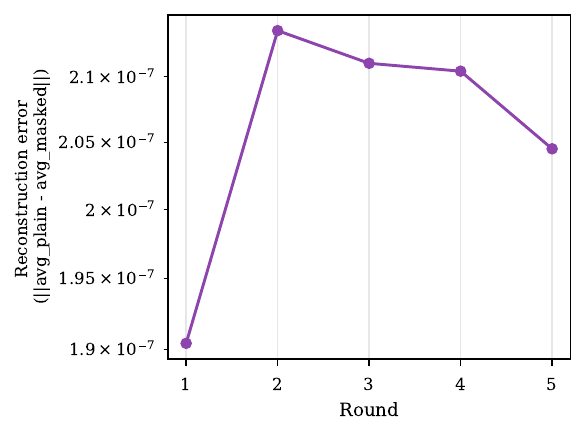}
        \caption{Channel estimation: reconstruction error (plain vs masked average) over rounds.}
        \label{fig:ce_recon}
    \end{minipage}\hfill
    \begin{minipage}{0.40\linewidth}
        \centering
        \includegraphics[width=\linewidth]{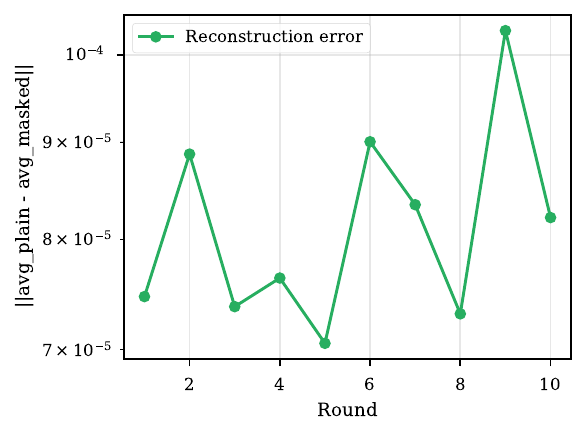}
        \caption{Radar sensing: reconstruction error (plain vs masked average) over rounds.}
        \label{fig:rad_recon}
    \end{minipage}
\end{figure*}

\textbf{Reconstruction error:} We measure $\|\bar{\boldsymbol{\theta}}_{\mathrm{ plain}} - \bar{\boldsymbol{\theta}}_{\mathrm{masked}}\|$ over rounds, where $\bar{\boldsymbol{\theta}}_{\mathrm{plain}}$ is the average of unmasked local models and $\bar{\boldsymbol{\theta}}_{\mathrm{masked}}$ is the average of masked updates. In theory, these should coincide; in practice, floating-point effects yield reconstruction errors on the order of $10^{-7}$ (channel) and $10^{-5}$ (radar). Figures~\ref{fig:ce_recon} and~\ref{fig:rad_recon} confirm that reconstruction error remains negligible across rounds.

\textbf{Leakage proxies:} We compute cosine similarity and Pearson correlation between the true local update $\boldsymbol{\theta}^{(k)} - \boldsymbol{\theta}^{\mathrm{global}}$ and the masked delta $\tilde{\boldsymbol{\theta}}^{(k)} - \boldsymbol{\theta}^{\mathrm{global}}$. High correlation would indicate information leakage. Table~\ref{tab:leakage} reports round-wise leakage proxies for channel estimation (secure mode). Correlations decrease over rounds as the masked perturbation diverges from the true update, suggesting effective obfuscation.

\begin{table}[t]
\centering
\caption{Channel Estimation---Leakage Proxies (Secure Mode, Sample Rounds)}
\label{tab:leakage}
\begin{tabular}{ccccc}
\toprule
Round & NMSE & QBER & Cosine $r$ & Pearson $r$ \\
\midrule
0 & 0.2261 & 0 & 0.945 & 0.943 \\
2 & 0.2170 & 0 & 0.884 & 0.877 \\
4 & 0.2156 & 0 & 0.840 & 0.828 \\
\bottomrule
\end{tabular}
\end{table}

\begin{figure*}[htbp]
    \centering
    \begin{minipage}{0.40\linewidth}
        \centering
        \includegraphics[width=\linewidth]{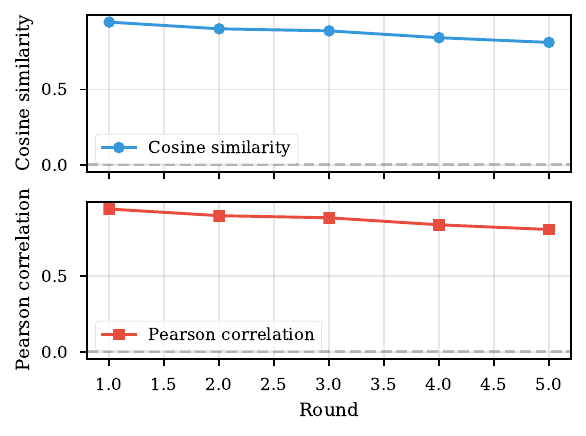}
        \caption{Channel estimation: cosine and Pearson correlation between true and masked updates.}
        \label{fig:ce_leakage}
    \end{minipage}\hfill
    \begin{minipage}{0.40\linewidth}
        \centering
        \includegraphics[width=\linewidth]{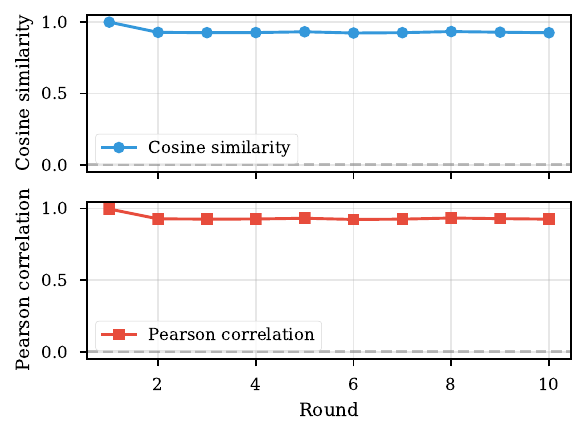}
        \caption{Radar sensing: cosine and Pearson correlation between true and masked updates.}
        \label{fig:rad_leakage}
    \end{minipage}
\end{figure*}

Figures~\ref{fig:ce_leakage} and~\ref{fig:rad_leakage} show leakage proxy evolution. The secure aggregation design ensures that individual client updates are not recoverable from the masked values alone.




\section{Discussion}
\label{subsec:discussion}

The experiments demonstrate that QKD-secured federated learning achieves near-parity with baseline FL in terms of utility, while providing strong security guarantees against eavesdropping. For channel estimation, QKD-SA and plain FL converge to nearly identical NMSE across all tested client counts $K \in \{3,10,20\}$ (e.g., $5.1\times 10^{-2}$ vs.\ $5.0\times 10^{-2}$ at $K=3$), indicating that pairwise masking does not introduce aggregation bias in practice. For radar sensing, QKD-SA attains slightly higher accuracy and mIoU than the plain baseline at $K=3$ and remains competitive for larger $K$, even in the presence of non-iid client data.

The threat model evaluation (Experiment B) confirms that QBER-based abort effectively prevents aggregation under attack. When an eavesdropper (Eve) performs an intercept-resend attack in every round, QBER rises to approximately $23$--$26\%$, exceeding the threshold $\tau_{\mathrm{QBER}}$. All rounds are aborted, and the model retains only the prior global state---accuracy remains at the random-guessing level ($6.96\%$) for radar sensing. This behavior is by design: the protocol refuses to aggregate when keys may be compromised, ensuring that an attacker cannot induce the server to use corrupted keys.

Experiment C characterizes the operational envelope of the system. As depolarizing channel noise $\eta$ increases, QBER rises monotonically. The abort threshold provides a clear security boundary: for $\eta$ low enough that QBER stays below $\tau_{\mathrm{QBER}}$, aggregation proceeds; above that point, rounds abort. This allows system operators to tune the threshold based on expected channel quality and desired security level.

Reconstruction error remains negligible (${\sim}10^{-7}$ for channel estimation, ${\sim}10^{-5}$ for radar), confirming that the plain and masked averages align within floating-point precision. Leakage proxies (cosine similarity and Pearson correlation between true and masked updates) decrease over rounds---from $0.94$ at round $0$ to $0.83$ at round $4$ for channel estimation---suggesting that the masked perturbation increasingly diverges from the true update. While nonzero correlation indicates some residual structure, the server and eavesdroppers observe only masked values; recovering individual client updates from the aggregate is infeasible without the pairwise keys.

Several limitations merit mention. Future work includes integration with measurement-device-independent QKD (MDI-QKD) \cite{PhysRevLett.108.130503}, which removes detector side-channel assumptions. The experiments use a protocol-level QKD abstraction rather than physical-layer MDI-QKD; real deployments would need to account for fiber or free-space quantum channels and their noise characteristics. The radar dataset is modest in size ($200$ samples); larger-scale evaluations would strengthen the conclusions. Furthermore, the leakage proxies are heuristic; formal privacy analysis (e.g., differential privacy) would require additional mechanisms. Future work may combine QKD-seeded masking with verifiable secure aggregation schemes to mitigate Byzantine updates \cite{zhou2025group,dong2023privacy}. Despite these caveats, the results support the feasibility of QKD-inspired security for federated learning in wireless and sensing applications, where joint radar--communication operations in 6G networks require both utility and protection against eavesdropping.


\section{Conclusion}
\label{sec:conclusion}

This paper proposes a QKD-secured federated learning framework for channel estimation and radar spectrum sensing in 6G networks. Using a BB84-style protocol abstraction and pairwise additive masking, clients train local models (CNN for channel estimation, U-Net for radar segmentation) and upload only masked updates. The server aggregates without observing plain parameters; eavesdroppers without QKD keys cannot recover individual contributions.

Experiments demonstrated that secure FL achieves utility parity with baseline FL: NMSE of $0.2156$ for channel estimation (matching baseline $0.2162$) and $92.13\%$ accuracy with $0.72$ mIoU for radar sensing. When an eavesdropper performs an intercept-resend attack, QBER rises to $\sim$25\% and all rounds abort as designed; model quality remains at the random-guessing level. QBER sensitivity to depolarizing noise provides a clear security boundary, and reconstruction error stays negligible (${\sim}10^{-7}$ and ${\sim}10^{-5}$), confirming correct aggregation. Leakage proxies decrease over rounds, indicating effective obfuscation of individual updates.

Future work includes integration with physical-layer MDI-QKD over fiber or free-space quantum channels, larger-scale evaluations on extended radar datasets, and formal privacy analysis (e.g., differential privacy) to complement the heuristic leakage proxies. The results support the feasibility of QKD-inspired security for federated learning in wireless and sensing applications, where 6G networks require both model utility and protection against eavesdropping.


\bibliographystyle{IEEEtran}
\bibliography{refs}

@inproceedings{jain2025recent,
  title={Recent Advances in Next Generation Cellular Mobile Networks-5G, 5G-Adv., and 6G},
  author={Jain, Prem Chand},
  booktitle={2025 International Conference on Innovation in Computing and Engineering (ICE)},
  pages={1--6},
  year={2025},
  organization={IEEE}
}

@article{ogenyi2025comprehensive,
  title={A comprehensive review of AI-native 6G: integrating semantic communications, reconfigurable intelligent surfaces, and edge intelligence for next-generation connectivity},
  author={Ogenyi, Fabian Chukwudi and Ugwu, Chinyere Nneoma and Ugwu, Okechukwu Paul-Chima},
  journal={Frontiers in Communications and Networks},
  volume={6},
  pages={1655410},
  year={2025},
  publisher={Frontiers Media SA}
}

@book{kuzlu2025ai,
  title={AI in Next Generation Networks (5G and Beyond): Fundamentals, Security and Applications},
  author={Kuzlu, Murat and Catak, Ferhat Ozgur and Zhao, Yanxiao and Ozdemir, Gokcen},
  year={2025},
  publisher={Independently Published}
}

@inproceedings{bonawitz2017practical,
  author    = {Bonawitz, K. and et al.},
  title     = {Practical Secure Aggregation for Privacy-Preserving Machine Learning},
  booktitle = {Proceedings of the ACM Conference on Computer and Communications Security (CCS)},
  year      = {2017},
  pages     = {1175--1191}
}

@article{abasi20256g,
  title={6g mmwave security advancements through federated learning and differential privacy},
  author={Abasi, Ammar Kamal and Aloqaily, Moayad and Guizani, Mohsen},
  journal={IEEE Transactions on Network and Service Management},
  volume={22},
  number={2},
  pages={1911--1928},
  year={2025},
  publisher={IEEE}
}

@article{kumar2025brief,
  title={A brief review on quantum key distribution protocols},
  author={Kumar, Mandeep and Mondal, Bhaskar},
  journal={Multimedia Tools and Applications},
  volume={84},
  number={27},
  pages={33267--33306},
  year={2025},
  publisher={Springer}
}

@article{kaur2025applications,
  title={Applications of Machine Learning in Strengthening 6G Security},
  author={Kaur, Gaganjot and Shrivastava, Vineet and Khan, Surbhi Bhatia and Singh, Anshu},
  journal={Security and Privacy in 6G Communication Technology},
  pages={167--190},
  year={2025},
  publisher={Wiley Online Library}
}

@phdthesis{hou2025towards,
  title={Towards Trustworthy Intelligence: Safeguarding Privacy and Security in Machine Learning},
  author={Hou, Sizai},
  year={2025},
  school={Hong Kong University of Science and Technology (Hong Kong)}
}

@article{jain2025toward,
  title={Toward Smart 5G and 6G: Standardization of AI-Native Network Architectures and Semantic Communication Protocols},
  author={Jain, Kurunandan and Krishnan, Prabhakar and Pachiyannan, Prabu and Jaganathan, Logeshwaran and Khan, Muhammad Attique and Li, Yang},
  journal={IEEE Communications Standards Magazine},
  year={2025},
  publisher={IEEE}
}

@ARTICLE{11389802,
  author={Alwis, Chamitha De and Aouedi, Ons and Xu, Jiaming and Wang, Shen and Siriwardhana, Yushan and Hewa, Tharaka and Zeydan, Engin and Sandeepa, Chamara and Liyanage, Madhusanka},
  journal={IEEE Communications Surveys \& Tutorials}, 
  title={Federated Learning for 6G Security: A Survey on Threats, Solutions and Research Directions}, 
  year={2026},
  volume={},
  number={},
  pages={1-1},
  doi={10.1109/COMST.2026.3663434}}

@article{othman2025key,
  title={Key enabling technologies for 6G: The role of UAVs, terahertz communication, and intelligent reconfigurable surfaces in shaping the future of wireless networks},
  author={Othman, Wagdy M and Ateya, Abdelhamied A and Nasr, Mohamed E and Muthanna, Ammar and ElAffendi, Mohammed and Koucheryavy, Andrey and Hamdi, Azhar A},
  journal={Journal of Sensor and Actuator Networks},
  volume={14},
  number={2},
  pages={30},
  year={2025},
  publisher={MDPI}
}

@article{abba2025iot,
  title={IoT-5G and B5G/6G resource allocation and network slicing orchestration using learning algorithms},
  author={Abba Ari, Ado Adamou and Samafou, Faustin and Ndam Njoya, Arouna and Djedouboum, Assid{\'e} Christian and Aboubakar, Moussa and Mohamadou, Alidou},
  journal={IET Networks},
  volume={14},
  number={1},
  pages={e70002},
  year={2025},
  publisher={Wiley Online Library}
}

@article{catak2022security,
  title={Security hardening of intelligent reflecting surfaces against adversarial machine learning attacks},
  author={Catak, Ferhat Ozgur and Kuzlu, Murat and Tang, Haolin and Catak, Evren and Zhao, Yanxiao},
  journal={IEEE Access},
  volume={10},
  pages={100267--100275},
  year={2022},
  publisher={IEEE}
}

@inproceedings{bb84,
  author    = {Bennett, Charles H. and Brassard, Gilles},
  title     = {Quantum Cryptography: Public Key Distribution and Coin Tossing},
  booktitle = {Proceedings of the IEEE International Conference on Computers, Systems and Signal Processing},
  year      = {1984},
  pages     = {175--179}
}

@inproceedings{sun2023massive,
  title={Massive mimo},
  author={Sun, Haijian and Ng, Chris and Huo, Yiming and Hu, Rose Qingyang and Wang, Ning and Chen, Chi-Ming and Vasudevan, Kasturi and Yang, Jin and Montlouis, Webert and Ayanda, Dauda and others},
  booktitle={2023 IEEE Future Networks World Forum (FNWF)},
  pages={1--70},
  year={2023},
  organization={IEEE}
}

@inproceedings{catak2024spectrumm,
  author    = {Catak, F. O. and Kuzlu, M.},
  title     = {A Federated Adversarial Learning Approach for Robust Spectrum Sensing},
  booktitle = {Proceedings of the Mediterranean Conference on Embedded Computing (MECO)},
  year      = {2024},
  pages     = {1--4}
}

@inproceedings{catak2023crypto,
  author    = {Catak, F.O. and Kuzlu, M.},
  title     = {A Cryptographic Federated Learning-Based Channel Estimation for Next-Generation Networks},
  booktitle = {Proceedings of the IEEE Virtual Conference on Communications (VCC)},
  year      = {2023},
  pages     = {92--97}
}

@article{seo2026gc,
  title={GC-Fed: Gradient Centralized Federated Learning with Partial Client Participation},
  author={Seo, Jungwon and Catak, Ferhat Ozgur and Rong, Chunming and Hong, Kibeom and Kim, Minhoe},
  journal={Information Fusion},
  pages={104148},
  year={2026},
  publisher={Elsevier}
}

@article{PhysRevLett.85.441,
  title = {Simple Proof of Security of the BB84 Quantum Key Distribution Protocol},
  author = {Shor, Peter W. and Preskill, John},
  journal = {Phys. Rev. Lett.},
  volume = {85},
  issue = {2},
  pages = {441--444},
  numpages = {0},
  year = {2000},
  month = {Jul},
  publisher = {American Physical Society},
  doi = {10.1103/PhysRevLett.85.441},
  url = {https://link.aps.org/doi/10.1103/PhysRevLett.85.441}
}

@article{PhysRevLett.108.130503,
  title = {Measurement-Device-Independent Quantum Key Distribution},
  author = {Lo, Hoi-Kwong and Curty, Marcos and Qi, Bing},
  journal = {Phys. Rev. Lett.},
  volume = {108},
  issue = {13},
  pages = {130503},
  numpages = {5},
  year = {2012},
  month = {Mar},
  publisher = {American Physical Society},
  doi = {10.1103/PhysRevLett.108.130503},
  url = {https://link.aps.org/doi/10.1103/PhysRevLett.108.130503}
}

@article{zhou2025group,
  title={Group verifiable secure aggregate federated learning based on secret sharing},
  author={Zhou, Sufang and Wang, Lin and Chen, Liangyi and Wang, Yifeng and Yuan, Ke},
  journal={Scientific Reports},
  volume={15},
  number={1},
  pages={9712},
  year={2025},
  publisher={Nature Publishing Group UK London}
}

@article{dong2023privacy,
  title={Privacy-preserving and Byzantine-robust federated learning},
  author={Dong, Caiqin and Weng, Jian and Li, Ming and Liu, Jia-Nan and Liu, Zhiquan and Cheng, Yudan and Yu, Shui},
  journal={IEEE Transactions on Dependable and Secure Computing},
  volume={21},
  number={2},
  pages={889--904},
  year={2023},
  publisher={IEEE}
}








\end{document}